\documentclass[10pt,a4paper,twocolumn,showpacs,nofootinbib,floatfix]{revtex4}
\usepackage{epsfig}
\usepackage{amsmath}
\usepackage{amsfonts}
\usepackage{amsbsy}
\usepackage{graphicx}
\renewcommand{\vec}[1]{\mathbf{#1}}
\def\be{\begin{equation}}
\def\ee{\end{equation}}
\def\bea{\begin{eqnarray}}
\def\eea{\end{eqnarray}}

\begin{document}
\title{Covariant description of kinetic freeze out through a finite time-like layer}


\author{
E. Moln\'ar$^{1,2}$, L. P. Csernai$^{1,3}$, V. K. Magas$^{1,4}$, Zs. I. L\'az\'ar$^{1,5}$,
A. Ny\'iri$^{1,6}$ and K. Tamosiunas$^1$ }

\bigskip

\affiliation{$^1$ Theoretical and Energy Physics Unit, Institute for Physics and Technology,
University of Bergen, Allegaten 55, N-5007 Bergen, Norway \\
$^2$ Frankfurt Institute for Advanced Studies, J. W. Goethe University, Max-von-Laue-Str. 1,
D-60438 Frankfurt am Main, Germany \\
$^3$ MTA-KFKI, Research Inst of Particle and Nuclear Physics,
H-1525 Budapest 114, P.O.Box 49, Hungary\\
$4$ Departament d'Estructura i Constituents de la Mat\'eria,
Universitat de Barcelona, Diagonal 647, 08028 Barcelona, Spain \\
$^5$ Department of Theoretical and Computational Physics,
Faculty of Physics, Babe\c s-Bolyai University, Str. M.
Kog$\breve{a}$lnicenau, nr. 1B, 400084 Cluj-Napoca, Romania \\
$^6$ Department of Physics, P.O.Box 1048 Blindern, N-0316, University of Oslo, Norway }

\begin{abstract}
{The Freeze Out (FO) problem is addressed for a covariant FO probability and a finite FO layer
with a time-like normal vector continuing the line of studies introduced in Ref. \cite{article_1}.
The resulting post FO momentum distribution functions are presented and discussed.
We show that in general the post FO distributions are non-thermal and asymmetric distributions
even for time-like FO situations.}
\end{abstract}


\maketitle
\section{Introduction}

Freeze out (FO) is a term referring to the stage of expanding or exploding matter when
its constituents (particles) loose contact, collisions cease and the local dynamical
equilibrium no longer can be maintained.
When the local equilibrium is significantly perturbed, the microscopic length and time scales
become comparable to the characteristic macroscopic ones and the hydrodynamical approach used
to describe the evolution of matter breaks down.
In the absence of collisions the momentum distribution of the particles "freezes out", hence
the name kinetic freeze out.
\\ \indent
The final break-up corresponds to a "phase-transition" from an interacting fluid to a non-interacting
gas of particles, where the interactions between the constituents ceases suddenly when reaching
the "critical" FO temperature of the order of the pion mass
$T_{FO}\approx 140$ MeV, as first assumed by Landau \cite{Landau_1}.
Consequently, FO is a discontinuity in space-time represented by a space-time boundary or
FO hypersurface, taken at the critical temperature [i.e., FO isotherm].
Across such FO hypersurface the properties of matter change suddenly.
We denote the two sides of the FO hypersurface as Pre FO and Post FO sides.
Originally, the Post FO distribution function was assumed to be an equilibrated J\"uttner distribution
function boosted with the local flow velocity on the actual side of the FO hypersurface.
This approximation and method corresponds to the so-called "sudden" FO model
by Cooper and Frye \cite{Cooper-Frye}.
\\ \indent
The Cooper-Frye type of FO process is the zero thickness limit of a more realistic, so-called
"gradual" FO process, where the FO description applies over a finite space-time domain, i.e. FO layer.
Inside the finite FO layer the properties of the matter change gradually trough interactions, while the
frozen out particles are formed and emitted at different "temperatures" which correspond to the
actual temperature of the interacting matter, gradually during the whole evolution of the matter.
\\ \indent
The basic philosophy of this paper is similar to the recent work \cite{article_1}, which introduces
and analyzes in detail the gradual FO description for space-like FO situations.
Through the paper we are going to use the notation from Ref. \cite{article_1}, recall
the governing equations and the major results for comparison.
We have made this paper sole and complete and widely understandable without the need to consult
our previous work where the original ideas were first introduced.
\\ \indent
Time-like discontinuities represent the overall sudden change in a finite volume where the events happen
simultaneously at causally disconnected points of the hypersurface with a time-like normal vector.
For example the assumption of instantaneous or isochronous FO [i.e., happening at a constant time in
the center of mass system], belongs to this category.
\\ \indent
The aim of this paper is to present an analyze a simple gradual kinetic FO process with time-like normal vectors.
In the first part of this study we introduce and generalize the gradual kinetic FO treatment for a finite
time-like FO layer in a fully covariant footing, while in the second part we analyze its outcome.

\section{Freeze-out from a finite time-like layer}\label{Fid}

\begin{figure}[t!]
\centering
\includegraphics[width=8.5cm, height = 8.2cm]{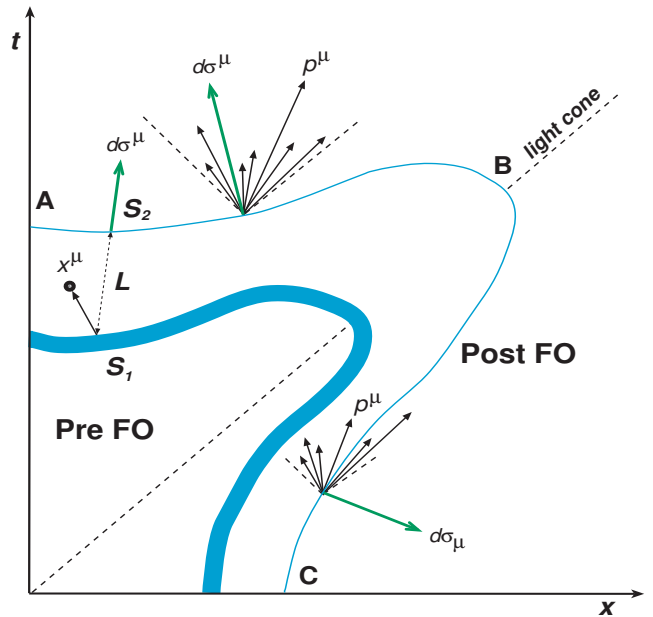}
\caption{(Color online) The figure shows a finite FO layer with varying thickness.
The normal to a surface element is $d\sigma_{\mu}$, thus between A and B the surface is
time-like, [i.e., $d\sigma^{\mu} d\sigma_{\mu} = + 1$],
from B down to C it is space-like, [i.e., $d\sigma^{\mu} d\sigma_{\mu} = -1$].
The change from time-like to space-like surface happens where the normal of the surface is light-like,
but not necessarily has its origin at the center of the system.
The momentum of particles is $p^{\mu}$ and the four vector $x^{\mu}$ denotes the particle coordinate,
having its origin at the inner surface of the FO layer.
On the time-like FO region all particles emerging from a point on the Pre FO side will propagate
to the Post FO side, while the particles originating from the space-like part of the FO surface
are divided between Pre FO and Post FO parts.
Only those particles cross the surface which have their momentum enclosed by the light cone and
the Post FO surface, [i.e., if $p^{\mu}d\sigma_{\mu}>0$].}
\label{figure_1t}
\end{figure}

The basis of the gradual FO method is to separate the "full" $f=f(x,p)$ distribution function into still
interacting and already frozen out parts, $f=f^i + f^f$, and describe the evolution of both components
in a self consistent way \cite{grassi_1, cikk_1, cikk_2, cikk_3, hama_1}.
This can be achieved by introducing the so-called escape rate, $\mathcal{P}_{esc}(s,p)$, used
to drain particles, which no longer collide from the interacting component, $f^i$, and to
gradually build up the free component, $f^f$.
For the better understanding of the model we use Fig. \ref{figure_1t}, and assume that the FO
of particles starts from the inside boundary of the FO layer, $S_1$ (thick line).
Within the FO layer of finite thickness, $L$, the density of interacting particles
decreases and disappears once we reach the outside boundary, $S_2$ (thin line),
of the FO layer.
\\ \indent
In general, the kinetic description of freeze out leads to a complex multidimensional problem.
To clarify the basic properties of the FO process through a finite layer we may essentially
reduce the number of variables, assuming that the dominant change in the distribution function
happens in the direction of the FO normal vector, while it is negligible along the directions
perpendicular to it, (e.g. in a spherically symmetric system the change happens in radial direction,
and it is negligible in the perpendicular directions).
Thus, the FO process can be effectively described as a one-dimensional process and the space-time domain where such a process takes place can be viewed as a FO layer, where the FO normal vector
is tied to the direction of the density decrease arising from velocity divergence at a curved FO surface.
\\ \indent
If we have a space-like normal vector, $d\sigma_{\mu}=(0,1,0,0)$, the resulting equations can be transformed
into a frame where the process is stationary, while in the case of a time-like normal vector,
$d\sigma_{\mu}=(1,0,0,0)$, the equations can be transformed into a frame where the process is
uniform and time-dependent.
In this paper we only discuss FO processes inside a finite time-like layer.
\\ \indent
Here we recall the governing equations [i.e., Eqs. (13-14)] from Ref. \cite{article_1},
which can be used in both time-like and space-like FO cases.
The equations depend on the projection, $s = x^{\mu} d\sigma_{\mu}$, in the direction of the FO normal
vector, $d\sigma_{\mu}$, where the four vector $x^{\mu}$ denotes the particle coordinate,
having its origin at the inner surface of the FO layer.
Thus,
\bea\label{first} \nonumber
\partial_s f^{i} (s,p) \! &=& \! - \mathcal{P}_{esc}(s,p) \, f^{i}(s,p) + \frac{f^{i}_{eq}(s,p) - f^{i}(s,p)}{\lambda_{th}} \, , \\
\partial_s f^{f} (s,p) \! &=& \! + \mathcal{P}_{esc}(s,p) \, f^{i}(s,p) \, ,
\eea
where using the relaxation time approximation we ensure that the interacting component approaches
the equilibrated J\"uttner distribution, $f^i_{eq}$, with $\lambda_{th}$ relaxation time
(or relaxation length in the space-like case).
\\ \indent
The escape rate, $\mathcal{P}_{esc}(s,p)$, describes the escape of particles from the interacting component
into the free component and it is defined as:
\be\label{esc2}
\mathcal{P}_{esc}(s,p) = \frac{1}{\lambda(s)}
\left[ \frac{L}{L - s} \, \frac {p^\mu d\sigma_\mu}{p^\mu u_\mu}\right] \Theta(p^{\mu}d\sigma_{\mu}) \, ,
\ee
where the parameter, $L$, is the "proper" thickness of the FO layer and it is an invariant scalar.
The proper thickness is analogous to the proper time, that is the time measured in the rest
frame of the particle.
In our case the local rest frame is the rest frame attached to the FO front (RFF), (see \ref{frames}),
and thus the proper "thickness" of the FO layer is the invariant proper time interval between the start
of the process and its end, (see Fig. \ref{figure_1t} at point A).
Furthermore, $p^{\mu}$ is the four-momentum of particles, $d\sigma_{\mu}$ is the normal in the FO direction,
while $u^{\mu}$ is the flow velocity normalized to unity.
The initial characteristic time is denoted by $\tau_0$, and the $\Theta(p^{\mu}d\sigma_{\mu})$ function,
was first introduced by Bugaev \cite{Bugaev_1} to ensure that all particles leave to the outside.
In case of time-like FO this condition is always satisfied and does not lead to any additional constraint.
\\ \indent
A qualitative expression of the escape rate for both time-like and space-like FO situations is based
on the following simple assumptions.
Particles with higher momentum in the FO direction will freeze out first.
The particles closer to the outside boundary of the FO layer have a greater chance
to freeze out since the probability to find another particle to collide with is smaller
as the system became sparser, (given by the $L/(L-s)$ factor).
In our special case the FO direction is parallel to the time-axis, thus all particles will freeze out
irrespective of their momenta.
However, particles emitted at later times will freeze out "faster" since they have less chance
to collide with other particles in the diluted system.
Please note that, although $\tau_0$ is assumed to be a constant for simplicity, the characteristic
FO length is increasing with time or distance such as, $\tau_0(L - s)/L$.
The detailed treatment and analysis of the escape rate can be found in \cite{article_1, ModifiedBTE_1, QM05_1}.
\\ \indent
Now, if we describe the time evolution of the particle FO, then $d\sigma_{\mu} = (1,0,0,0)$,
$x^{\mu}d\sigma_{\mu} = t$ and $p^{\mu}d\sigma_{\mu} = p^0$, thus eq. (\ref{first}) leads to
\bea\label{first-rethermalized} \nonumber
\partial_t f^{i} &=& - \frac{1}{\tau_0} \left( \frac{L}{L-t} \right)\!
\left( \frac{p^0}{p^{\mu} u_{\mu}} \right) f^{i}
+  \frac{ f^{i}_{eq} - f^{i}}{\tau_{th}} \, , \\
\partial_t f^{f} &=& + \frac{1}{\tau_0} \left(\frac{L}{L-t} \right)\!
\left(\frac{p^0}{p^{\mu} u_{\mu}}\right) f^{i} \, ,
\eea
where the interacting component approaches the equilibrated J\"uttner distribution, $f_{eq}$,
with $\tau_{th}$ relaxation time.
This is a common simplification and the practical reason to use it is to calculate the quantities
depending on the equilibrium distribution function.
Although the present solution mathematically is achieved taking an infinitely short relaxation time, in reality,
one can show, see Ref. \cite{article_1}, that the complete thermalization of the interacting component can
be achieved with good accuracy if $\tau_0$ is smaller than $\tau$ by a factor of 2 or more.
For a thorough analysis of this approach, see Refs. \cite{article_1, sven} and the Appendix.
Here we mention that in our calculations we use only one type of particles,
namely massless pions, therefore the chemical composition of our system remains unchanged during
the kinetic FO.
Furthermore, for simplicity we assume simultaneous chemical and thermal equilibration, thus the
chemical potential of the massless pion gas is $\mu=0$ during the FO process.
\\ \indent
Earlier in Ref. \cite{cikk_5}, the gradual time-like FO description was modeled with equations
having a similar form to eqs. (\ref{first-rethermalized}), but it was only treating the simplest
case when $u^{\mu}=d\sigma_{\mu}$, while the FO was lasting infinitely long.
The model was based on the idea of the boost invariant Bjorken hydrodynamical model \cite{Bjorken},
where the evolution of matter is a  function of the proper time, $\tau$, only, while the flow
of matter is parallel to the normal vector of the proper time hyperbolas in every point.
The covariant equations, eqs. (\ref{first-rethermalized}), return those equations if we change
the time variable, $dt= d\tau$, and neglect the $L/(L-t)$ factor.
However, the new equations allow $u^{\mu}\neq d\sigma_{\mu}$ and also a possibility to finish the FO process
within a given duration.

\subsection{Reference frames}
\label{frames}

Before proceeding further, first we define the reference frames in which our calculations
will be handled.
\begin{figure}[!hbt]
\centering
\includegraphics[width = 8.4cm, height = 7.2cm]{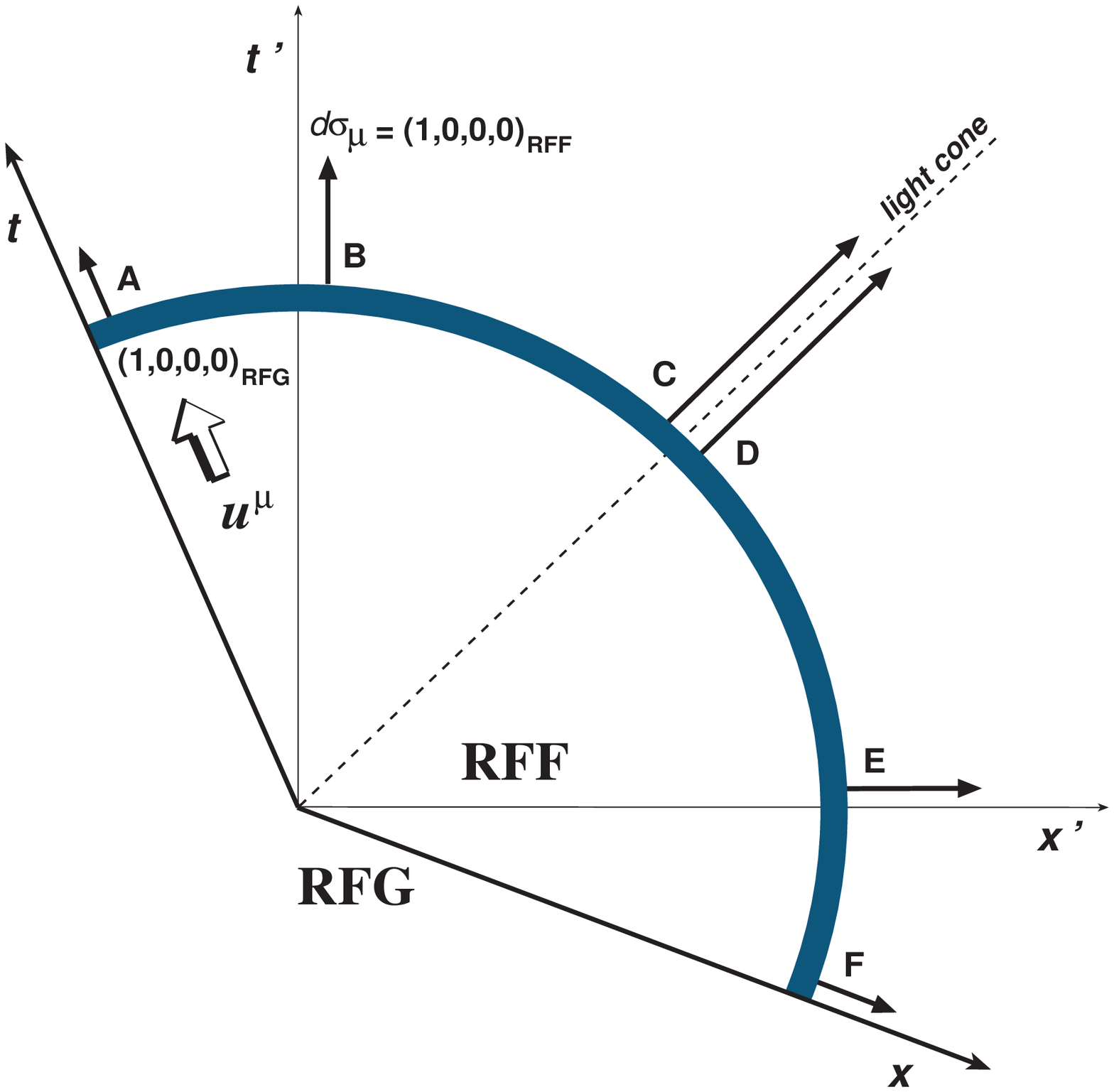}
\caption{(Color online) A simple FO hypersurface in RFG with coordinates [t,x],
where $u^{\mu} = (1,0,0,0)_{RFG}$.
The normal vectors of the FO front, $d\sigma_{\mu}$, are time-like at points, A, B, C,
while the normal vectors are space-like at points, D, E, F.
At point B in RFF with coordinates [t',x'], where $d\sigma_{\mu}=(1,0,0,0)_{RFF}$ and $u^{\mu} = \gamma_{\sigma}(1,-v_{\sigma},0,0)_{RFF}$.
Note that RFF moves together with the FO front, while on the figure the origin of RFF is shifted to match
the origin of RFG.
}
\label{figure_2t}
\end{figure}
\\ \indent
On the Pre FO side the matter is parameterized by an equilibrium distribution function,
as required by the hydrodynamical description of evolution.
The frame where the matter is at rest is the local Rest Frame of the Gas (RFG), where $u^{\mu} = (1,0,0,0)_{RFG}$.
On the Post FO side we can use the frame which is attached to the FO front, that is the
Rest Frame of the Front (RFF). In RFF the normal vector to the FO hypersurface for the time-like part is always
$d\sigma_{\mu} = (1,0,0,0)_{RFF}$,  while on the space-like part is always $d\sigma_{\mu} = (0,1,0,0)_{RFF}$.
\\ \indent
If we are in the RFG then the four-flow is always $u^{\mu} = (1,0,0,0)_{RFG}$.
If we take different characteristic points, for example points, A, B and C, on the FO hypersurface
then the normal vector is different at different points of the hypersurface in RFG, see Fig. \ref{figure_2t}.
To calculate the parameters of the normal vector, $d\sigma_{\mu}$, for different cases in the RFF
we make use of the Lorentz transformation.
\\ \indent
The normal vector of the time-like part of the FO hypersurface may be defined as the local $t'$-axis,
while the normal vector of the space-like part may be defined as the local $x'$-axis.
This defines the axes of RFF.

\subsection{Conservation laws}\label{conservation}

The change of conserved quantities caused by the particle transfer from the interacting matter
into the free matter can be obtained in terms of distribution function of the interacting matter
calculated from eqs. (\ref{first-rethermalized}) as:
\bea
dN^{\mu}_{i} (t) \! &=& \! dt \! \! \int \frac{d^3 p}{p^0} \, p^{\mu} \, \partial_{t} f^{i} \\ \nonumber
\! &=& \! -  \frac{dt}{\tau_0} \frac{L}{L-t} \int \frac{d^3 p}{p^0}\, p^{\mu}
\, \frac{p^{\rho} d\sigma_{\rho}}{p^{\rho}u_{\rho}} f^i_{eq}(t,p) \, ,
\eea
while the change in the energy-momentum as:
\bea
dT^{\mu\nu}_{i} (t) \! &=& \! dt \! \! \int \frac{d^3 p}{p^0} \, p^{\mu} p^{\nu} \, \partial_{t} f^{i} \\ \nonumber
\! &=& \! - \frac{dt}{\tau_0} \frac{L}{L-t} \int \frac{d^3 p}{p^0}\, p^{\mu} p^{\nu}
\, \frac{p^{\rho} d\sigma_{\rho}}{p^{\rho}u_{\rho}} f^i_{eq}(t,p)\, .
\eea
The equilibrium distribution function for massless baryonfree particles is:
\be
f^i_{eq}(t,p) = \frac{g}{(2\pi \hbar)^3} \,
\exp{\left[-\frac{{\gamma(p^0 - jup \cos{\theta}_{\vec{p}})}}{T}\right]} \, ,
\ee
where the four-momentum of particles is $p^{\mu} = (p^0,\vec{p})$, $p=|\vec{p}|$, $p^x = p \cos \theta_{\vec{p}}$,
the flow velocity of the interacting matter in RFF is $u^{\mu} = \gamma(1,v,0,0)_{RFF}$,
$\gamma = 1/\sqrt{1 - v^2}$, $u = |v|$, $j = \textrm{sign} (v)$,
and $g$ is the degeneracy of particles.
The results of the calculations in the RFF can be found in Appendix B.
\\ \indent
The change in energy density after a step $dt$ is
\be\label{energy_density}
d e_{i}(t) = u_{\mu,i}(t) \, dT^{\mu\nu}_{i}(t) \, u_{\nu,i}(t) \, ,
\ee
from which by using a simple EoS, $e = \sigma_{SB} T^4$, for a baryonfree massless gas
with $\sigma_{SB} = \frac{\pi^2}{10}$, we can calculate the change
in the temperature and Landau's flow velocity similarly to \cite{article_1, cikk_1, cikk_3}.
Thus,
\bea\label{landau}
d \ln T &=& \frac{\gamma^{2}}{4\sigma_{SB} T^{4}} \bigg[ dT^{00}_i -
2vdT^{0x}_i + v^{2} dT^{xx}_i \bigg] \, ,\\ \nonumber
d v &=&  \frac{3}{4 \sigma_{SB} T^{4}}\bigg[ -v dT^{00}_i + (1+v^{2}) dT^{0x}_i -v dT^{xx}_i \bigg]\,.
\eea
while in the massless limit the above equations lead to:
\bea \label{massless_landau}
d \ln T &=& - \frac{dt}{\tau_0} \left(\frac{L}{L-t} \right)\frac{3n\gamma}{4\sigma_{SB} T^{3}} \, ,\\ \nonumber
d v &=& - \frac{dt}{\tau_0} \left(\frac{L}{L-t} \right)\frac{3 n v}{4 \gamma \sigma_{SB} T^{3}} \, .
\eea
\section{Results and discussions}

In this section we will present the results for the Post FO distribution and the relevant quantities
calculated form this model.
We will present our results for two different cases, for infinite FO (I) and finite FO (F).
\begin{itemize}
\item [ I) \,]
The system is characterized by an infinitely long FO duration, [i.e., $L/(L-t) \rightarrow 1$ and $(t = 100 \tau_0)$,
where most of the matter is frozen out].
The results are shown on
Figs. \ref{figure_3t}, \ref{figure_5t}, \ref{figure_7t}, \ref{figure_8t}.
\item [ F) \,] A finite FO process happening in a finite FO layer, where
$(L=10 \tau_0)$.  The results are shown on
Figs. \ref{figure_4t}, \ref{figure_6t}, \ref{figure_7t}, \ref{figure_9t}, \ref{figure_10t}.
\end{itemize}

\subsection{The evolution of temperature of the interacting component}

\begin{figure}[t!]
\centering
\includegraphics[width=8.5cm, height = 5.5cm]{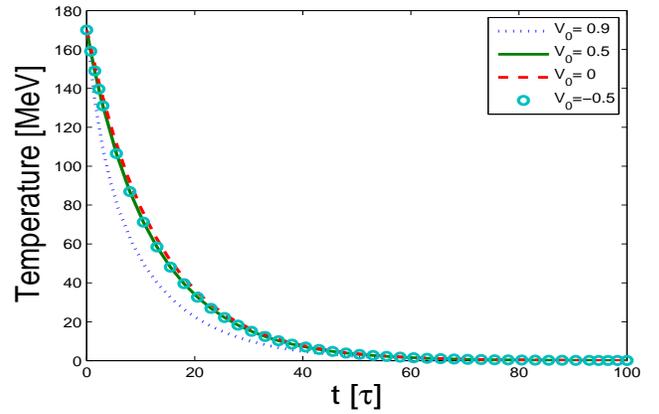}
\caption{(Color online) The temperature of the interacting component in RFF, calculated
for an infinitely lasting FO.
The initial temperature is \mbox{$T_0 = 170\,$ MeV}, the parameter, $v_0$, is the initial
flow velocity.
This corresponds to case I.}
\label{figure_3t}
\end{figure}

\begin{figure}[hbt!]
\centering
\includegraphics[width=8.5cm, height = 5.5cm]{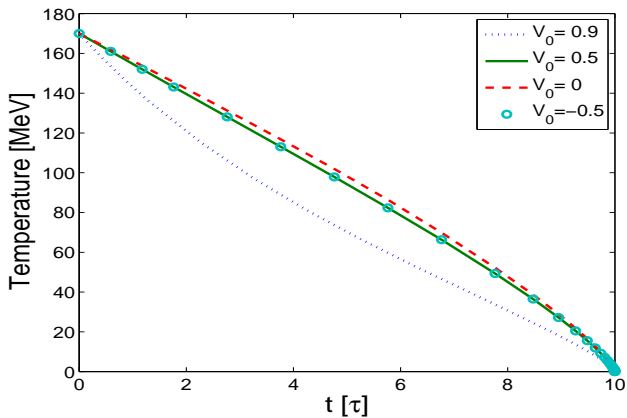}
\caption{(Color online) The temperature of the interacting component in RFF, calculated
for a finite $(L=10 \tau_0)$ FO time, where the initial temperature is \mbox{$T_0 = 170\,$ MeV},
and $v_0$ is the initial flow velocity.
This corresponds to case F.}
\label{figure_4t}
\end{figure}

The first set of figures, Fig. \ref{figure_3t} and Fig. \ref{figure_4t}, shows the gradual decrease in
temperature of the interacting component calculated in RFF.
\\ \indent
Comparing Fig. \ref{figure_3t} with Fig. \ref{figure_4t}, we see the difference between the
finite and infinite FO.
The FO in a finite layer is faster than in an infinite layer, "per se".
The temperature curves belonging to different initial flow velocities but with opposite
sign [i.e., $v_0 = -0.5$ and $v_0=0.5$] are the same.
This is so since for time-like FO we do not have any constraint on the momenta,
such as the cut-off factor $\Theta(p^{\mu}d\sigma_{\mu})$, and the initial momentum distribution
is symmetric over the time axis.
In the case of time-like FO the gradual cooling of the matter is faster and "smoother" compared
to space-like FO, since the most energetic particles freeze out unrestricted in direction
thus the remaining interacting component cools down faster.
\\ \indent
The matter with higher initial flow velocity, $v_0$, cools faster, but for small differences
between the initial flow velocities, the resulting difference in the temperature is negligible.
If the interacting gas has a higher flow velocity in RFF, the escape rate is bigger for
higher values of the flow velocity.
This expresses the fact that if the matter flows we remove energy faster form the
interacting component.
\\
\\

\subsection{The evolution of common flow velocity of the interacting component in RFF and RFG}

\begin{figure}[!t]
\centering
\includegraphics[width=8.5cm, height = 5.5cm]{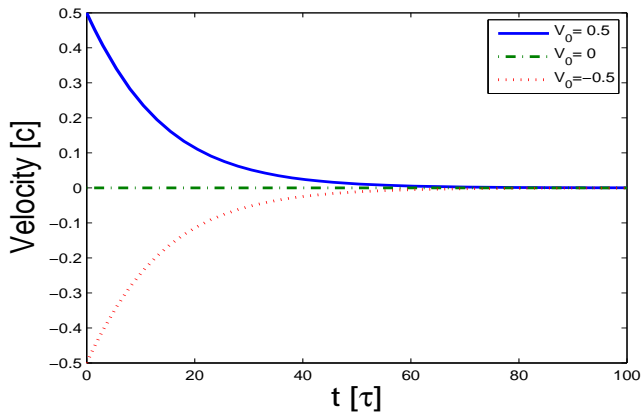}
\caption{(Color online) The evolution of the flow velocity of the interacting component calculated for an
infinitely lasting FO, corresponding to case I.
The initial temperature is $T_0 = 170\,$ MeV, and $v_0$ is the initial flow velocity of the gas.}
\label{figure_5t}
\end{figure}

\begin{figure}[!hbt]
\centering
\includegraphics[width=8.5cm, height = 5.5cm]{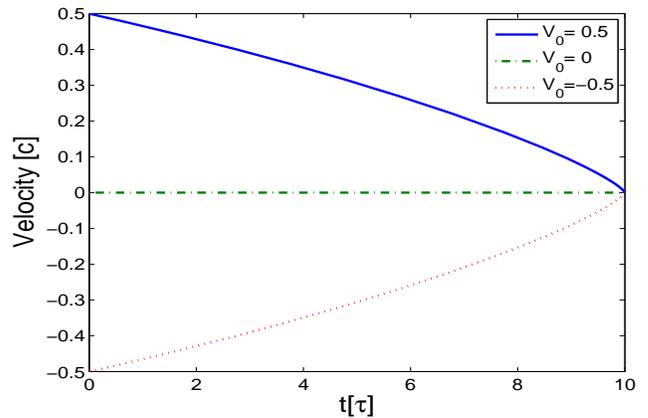}
\caption{(Color online) The evolution of the flow velocity of the interacting component calculated
for a finite $(L=10 \tau_0)$ FO, corresponding to case F.
The initial temperature is $T_0 = 170\, $ MeV, and $v_0$ is the initial
flow velocity of the gas.}
\label{figure_6t}
\end{figure}

The second set of figures, Figs. \ref{figure_5t}, \ref{figure_6t}, shows the evolution of
the flow velocity of the interacting component calculated for a baryonfree massless gas in RFF.
\\ \indent
Again, comparing Fig. \ref{figure_5t} with Fig. \ref{figure_6t}, we can see the difference between
finite and infinite FO.
In the case of finite FO the velocity decrease is much faster than in the case of infinite FO.
\\ \indent
Furthermore, we notice that the flow velocity of the interacting component tends to
zero, while here we recall to compare, that in case of space-like FO it tends to $-1$.
Again, this is due to the cut-off factor which retains particles propagating with negative momenta
in space-like directions.
In RFF the quantities change discontinuously at the light cone and that is why we
have different results for the final flow velocity comparing space-like and time-like cases.
However, in RFG all quantities are continuous when crossing the light cone, see Ref. \cite{article_1,QM05_1}.
\\

\subsection{The transverse momentum and the contour plots of the Post FO distribution}

\begin{figure}[!hbt]
\centering
\includegraphics[width=8.5cm, height = 5.5cm]{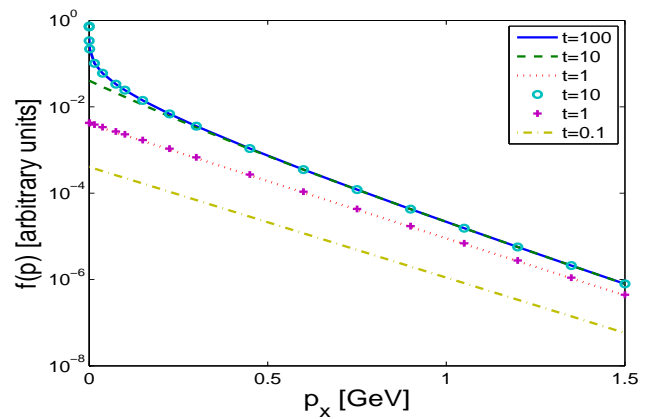}
\caption{(Color online) The local transverse momentum (here $p_x$) distribution for a baryonfree and massless
gas at $(p_y = 0)$.
The calculations were done for an infinite FO with thin lines and for
a finite FO $(L=10\tau_0)$ with marker lines.
The initial flow velocity and temperature are, $u^{\mu}=(1,0,0,0)_{RFG}$ and $T_0 = 170\,$MeV.
The transverse momentum spectrum at the end is obviously curved due to the FO process
for low momenta.}
\label{figure_7t}
\end{figure}

\begin{figure}[!t]
\centering
\includegraphics[width=8.5cm, height=3.8cm]{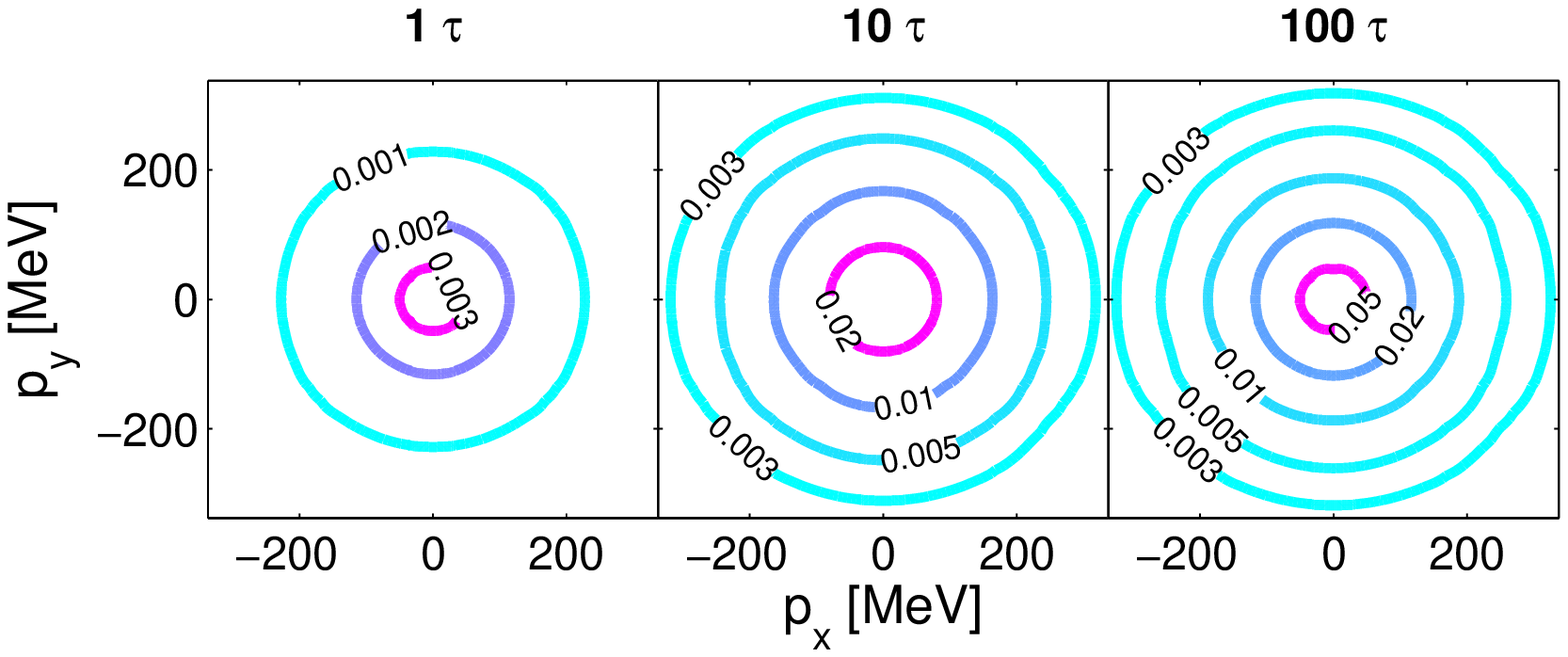}
\caption{(Color online) The Post FO distribution, $f_{free}(x,\vec{p})$, at point A of Fig. \ref{figure_2t}, for an infinitely
long FO length.
The figures correspond to different time points, {\bf $t = 1 \tau_0,\ 10 \tau_0,\  100 \tau_0$}
respectively.
Contour lines are given at values represented on the figure.
The initial flow velocity and temperature are; $u^{\mu}=(1,0,0,0)_{RFG}$ and $T_0 = 170\,$ MeV.
The maximum is increasing with $t$ as indicated on Fig. \ref{figure_7t}.
}
\label{figure_8t}
\end{figure}

The third set of figures, Figs. \ref{figure_7t}, and \ref{figure_8t}, shows the evolution of the local
transverse momentum distribution and the corresponding contour plots of the Post FO momentum distribution.
\\ \indent
We have presented a one-dimensional model here, but we assume that it is applicable for the direction
transverse to the beam in heavy ion experiments.
The plots presented should be related to the transverse momentum distribution of measured particles.
\\
\\ \indent
From Fig. \ref{figure_7t} we see that the Post FO momentum distributions for the
infinite and finite FO cases are qualitatively identical.
At the early stages of the FO process (for values of $t \simeq \tau_0$) the distribution
of particles in the two cases match.
This property persists until the end of the FO process, and can be seen on Fig. \ref{figure_7t}, where
the local transverse momentum distribution calculated for an infinite and finite FO are identical.
The maximum is increasing with $t$ as indicated in Figs. \ref{figure_7t} and \ref{figure_8t}.
Thus, the final Post FO distributions do not differ if we switch from an infinitely long to a finite
layer FO description for any initial flow velocity.
This means that our finite layer FO description was done correctly.
These important features of the model were already discussed in detail in Refs. \cite{article_1, QM05_2}.
\\ \indent
The overall conclusion is that the resulting Post FO distributions are non-thermal distributions even
for time-like FO processes.
The distributions strongly deviate from thermal distributions in the low momentum region [i.e., $p_x < 300$ MeV].
If one decreases the duration of total FO time, the gradual FO process would still produce similar
particle spectra until the duration is not less than $2\tau_0$.
Below that value the spectra becomes less curved and in the limit when the duration approaches zero
the final momentum spectra corresponds to a constant temperature equilibrium distribution function.
For $L < 2\tau_0$ the FO process does not have enough time to significantly change the shape of the
final spectrum.
\\ \indent
Here one may go further and intuitively say that the FO process has a maximal lifespan,
even though the parameter, L, was not defined in this work in such way that it would
allow us to exactly calculate its limits from first principles.
Of course for such a statement to hold one would need a realistic full scale fluid dynamical
simulation including chemistry, secondaries and the expansion of the system.
However, our results concluded from this simple model could still hold valuable in the realistic
FO modeling in complex fluid dynamical simulations.
\subsection{The boosted Post FO distributions}

\begin{figure}[!t]
\centering
\includegraphics[width=8.5cm, height =3.8cm]{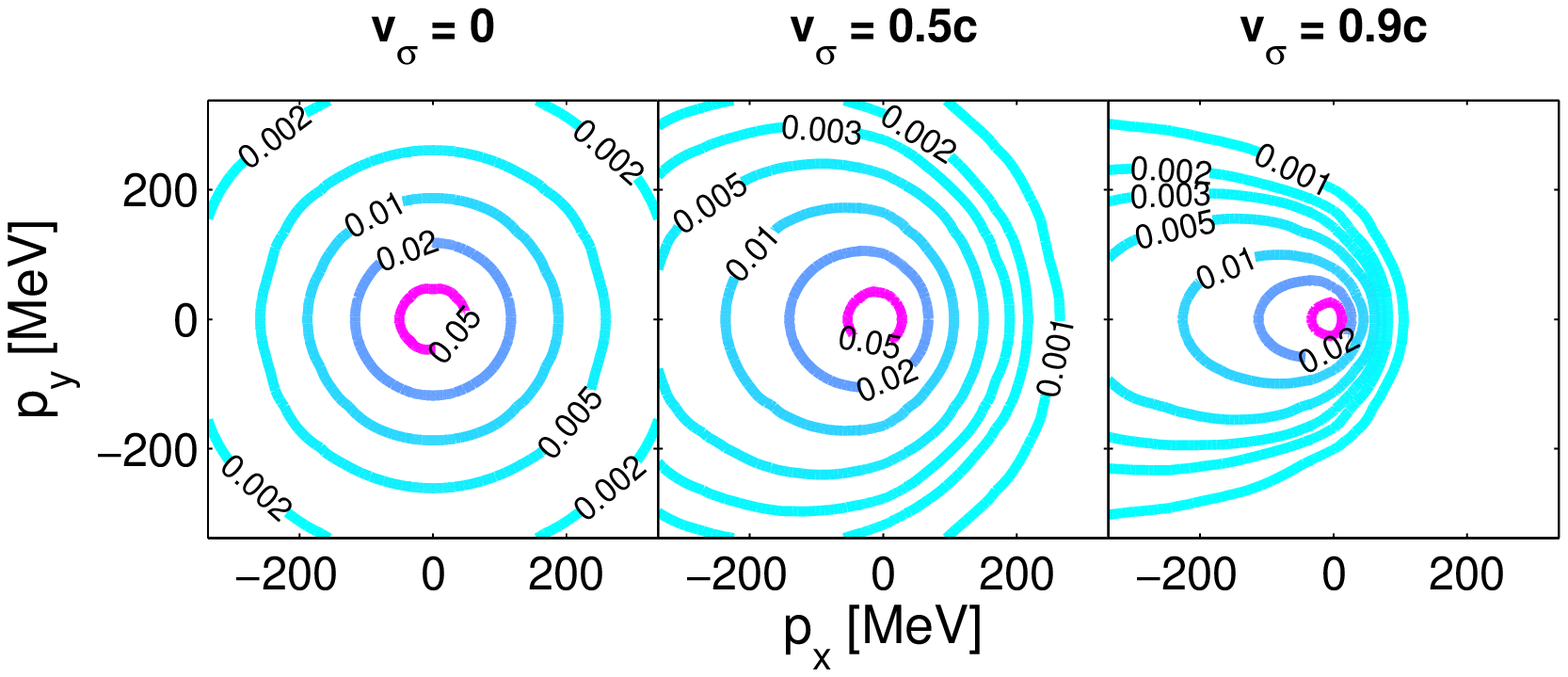}
\caption{(Color online) The final Post FO distribution, $f_{free}(x,\vec{p})$, at points A, B, C,  of Fig. \ref{figure_2t},
calculated for a finite FO time, $L = 10\tau_0$.
The contour plots correspond to different initial flow velocities,
$u^{\mu}=\gamma_{\sigma}(1,-v_{\sigma},0,0)_{RFF}$, with $v_{\sigma} = 0, 0.5c, 0.9c$, respectively
where the initial temperature is $T_0 = 170\, $ MeV.
Note that, the Post FO distributions at points B and C are not the boosted distributions of point A.
The difference is that the Post FO distribution is much less elongated than the boosted J\"uttner, because
it is a superposition of sources with decreasing speed in RFF as indicated in Fig. \ref{figure_6t}.}
\label{figure_9t}
\end{figure}

\begin{figure}[!t]
\centering
\includegraphics[width=8.5cm, height =3.8cm]{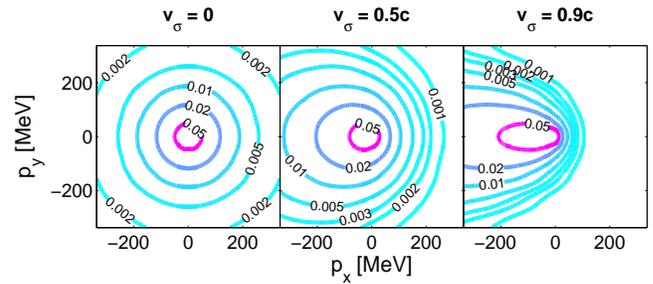}
\caption{(Color online) The post FO distribution at point A and boosted to the frames at points B and C as
depicted in Fig. \ref{figure_2t}.
The different figures correspond to different initial normal vectors,
$d\sigma^{\mu}=\gamma_{\sigma}(1,v_{\sigma},0,0)_{RFG}$,
where {\bf $v_{\sigma} = 0, 0.5c, 0.9c$}, and the initial temperature is $T_0 = 170\,$ MeV.}
\label{figure_10t}
\end{figure}

The fourth set of figures, Fig. \ref{figure_9t} and Fig. \ref{figure_10t}, shows the
final Post FO distributions calculated for different flow velocities in RFF and the
boosted post FO (J\"uttner) distributions in RFG.
The FO distributions corresponding to different initial flow velocities,
$u^{\mu}=\gamma_{\sigma}(1,-v_{\sigma},0,0)_{RFF}$, generally lead to
non-equilibrated and anisotropic Post FO distributions.
However, boosting the distribution from point A to points B or C leads to a more elongated FO distribution
in the direction of the boost than the calculated Post FO distributions at those points,
(compare the contours with the same values given on Figs.  \ref{figure_9t} and \ref{figure_10t}).
\\ \indent
This is also an important outcome of our analysis leading to the conclusion that assuming an
isotropic equilibrated J\"uttner distribution at time-like parts the FO hypersurface, other
than at point A on Fig. \ref{figure_2t} where $u^{\mu} = d\sigma_{\mu}$, in general cannot hold
\cite{gorenstein}.
More importantly, the common practice of boosting the J\"uttner distribution or any
post FO distribution function instead of calculating it from the conservation laws,
similarly as it was done here, leads to a noticeable difference in the final particle spectra.

\subsection{The non-equilibrated post FO distributions revised}

\begin{figure}[!hbt]
\centering
\includegraphics[width=8.5cm, height = 5.5cm]{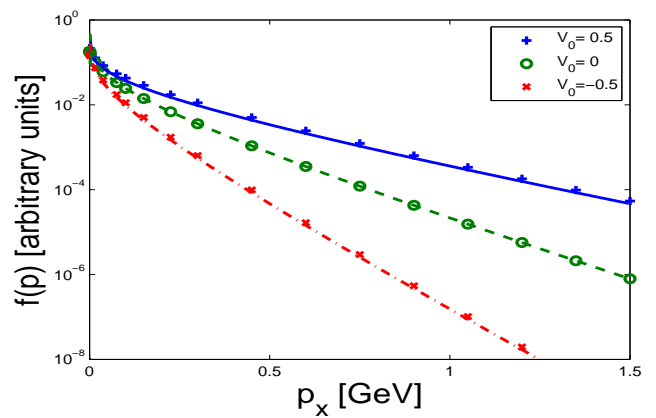}
\caption{(Color online) The local transverse momentum (here $p_x$) distribution for a baryonfree and massless
gas at $(p_y = 0)$.
The calculations were done for an infinite FO with thin lines using eq. (\ref{exp_int})
and for a finite FO $(L=3\tau_0)$ with marker lines.
The initial temperature was $T_0 = 170\,$MeV.
The different initial flow velocities are given on the figure legend.}
\label{figure_12t}
\end{figure}

Here we present another important result of our study, following the approach from
Ref. \cite{cikk_5}, where an infinitely long FO was studied with momentum independent escape rate.
We can reproduce that earlier result by taking $u^\mu(t_0)=d\sigma^\mu=(1,0,0,0)$, and can calculate the temperature decrease using eqs. (\ref{massless_landau}), in the case of a massless baryonfree matter for an
infinitely lasting FO:
\be
T(t) = T(t_0) \, \exp \left(-\frac{k}{\tau_0} (t-t_0) \right)\, ,
\ee
where $k = 3 /(4 \pi^2\sigma_{SB})$. In the general case, the flow velocity is not zero, hence
one has to solve the system of equations from eqs. (\ref{massless_landau}).
The distribution function of interacting particles (in the fast rethermalization limit) at any time $t$ is:
\be
f^i(t,p) = \frac{1}{(2\pi)^3} \, \exp \left(-\frac{p^0}{T(t_0)} \, e^{k (t-t_0)/\tau_0}\right) \, ,
\ee
Now, we can solve the equation for the free component from eqs. (\ref{first-rethermalized}), therefore the distribution of free particles at time $t$ is:
\bea\label{exp_int_t}
f^f(t,p) &=& \frac{1}{\tau_0}\int_{t_0}^{t} f^i(t',p) \\ \nonumber
&=&\frac{k^{-1}}{(2\pi^3)} \, \left[ \textrm{Ei} \left( -\frac{p^0}{T(t)} \right)
-  \textrm{Ei} \left( -\frac{p^0}{T(t_0)}\right)  \right]\, ,
\eea
which for $t\rightarrow \infty$ leads to:
\be\label{exp_int}
f^f = \frac{k^{-1}}{(2\pi^3)} \, \textrm{Ei} \left( -\frac{p^{\mu} u_{\mu}(t_0)}{T(t_0)}\right)
\ee
where $\textrm{Ei}$ is the exponential integral function defined in Appendix B eq. (\ref{exponential_integral}).
Thus, we got a simple formula, similarly to the one in Ref. \cite{cikk_5}, which correctly parameterizes
the non-equilibrated post FO distribution function when $u^{\mu} = (1,0,0,0)$.
It was actually shown in Ref. \cite{article_1} that the post FO distribution is not
sensitive to momentum dependence of the escape rate, so we assume that this simple formula
is valid for any initial flow velocity $u^\mu(t_0)$.
\\ \indent
Here we will use the above formula to plot the post FO distribution, for finite layers, with $L > 2\tau_0$,
and extend this approximation to the general case when $u^{\mu}(t_0)\neq d\sigma_{\mu}$.
On Fig. \ref{figure_12t}, we have plotted the final FO distribution functions calculated
using eq. (\ref{exp_int}) with lines, and the finite FO $(L=3\tau_0)$ calculation with marker lines.
The results are matching, which is a remarkable result, thus we conclude that this simple approximation
is applicable for the description of gradual FO thorough finite time-like layers and correctly
approximates its post FO distribution functions.

\section{Conclusions}

In this work we have presented a simple kinetic freeze out model for a finite time-like layer.
We have demonstrated that FO across time-like surfaces leads to non-equilibrated and anisotropic distributions.
These distributions in general cannot be Lorentz transformed to a frame where the distribution is isotropic.
The only exception is when the normal to the FO hypersurface is parallel to the local flow velocity.
Our analysis shows that the usual practice of assuming a J\"uttner distribution as a Post FO distribution is
in general not valid!
\\ \indent
We can also see that while the boosted J\"uttner distribution is elongated in the boost direction,
i.e. in the direction of $d\sigma_{\mu}$, the Post FO distribution is close to a spherical and isotropic
distribution at low momenta, and becomes elongated only at higher momenta, see Fig. \ref{figure_9t}.
This special Post FO distribution leads to a curved "$p_{t}$ - spectrum".
Here, we can also demonstrate (as in earlier works \cite{article_1, cikk_1, cikk_2, cikk_3}) that non-equilibrium
processes in kinetic FO lead to observable effects.
\\ \indent
We observe that the J\"uttner distribution is not a good approximation for the Post FO distribution,
just like in the case of a space-like FO.
While in the case of space-like FO the Cancelling-J\"uttner distribution introduced in Ref. \cite{karolis}
is satisfactory, in the case of time-like FO, we have found a simple formula to use.
\\ \indent
Now, one may ask the question whether we observe this additional low $p_t$ effect in the experimental data.
This effect has several possible explanations: products of low momentum resonance decays,
the transverse expansion of the system, and possibly due to the long gradual FO with rethermalization.
As already discussed, during such scenario the particles are freezing out at different
gradually decreasing temperatures, thus correspondingly the final FO spectrum is a
superposition of thermal distributions with different temperatures.
Although the low $p_t$ enhanced non-thermal spectrum of massless pions is a necessary outcome of long gradual
FO with rethermalization, for the heavy particles (if these are in the mixture with pions) it is almost
unobservable, see Ref. \cite{cikk_5}.
\\ \indent
In our simplistic study we have found that FO in layer of finite thickness below $L < 2\tau_0$ will
not show a sharp peak at low momentum in the transverse momentum spectrum.
Thus, naively one can conclude from a simple fit that the FO in heavy-ion collisions happens
in a narrow or wide FO layer.
However, such conclusion would be premature without including the expansion of the system and
calculate two particle correlations.
At the moment FO in a long finite layer, $L > 2\tau_0$, cannot be excluded.
If one assumes gradual FO with non-thermal post FO spectra, then one may also fit the data but with
different flow velocity and slope parameter, where the curvature of the pion spectra at low $p_t$ will
be partly due to FO and partly due to the expansion of the system.

\section{Outlook}

We do not aim directly to apply the results presented here to experimental heavy ion collision data,
instead our purpose was to study qualitatively the basic features of the freeze out process, and
to demonstrate the applicability of this covariant formulation for FO in a finite layer.
\\ \indent
Here we note that our model may be applicable in CFD calculations, where one has both time-like
and space-like parts of the full FO hypersurface, thus the gradual FO calculation must be done over
the full FO hypersurface, with varying flow velocities and normal vectors.
The method should be applied after reaching the $T_{FO}$ critical temperature and calculate the Post FO
momentum distribution function starting form the inner FO hypersurface.
Such a calculation should be compared to the Cooper-Frye ansatz in the first place and then to
experimental results, similarly to Refs. \cite{grassi_1, hama_2, grassi_2}.
\\ \indent
A successful application of this model was already used to study the impact of nucleon mass shift
on the freeze out process \cite{sven}.
This analysis will be carried forward to calculate the impact of mass shift on other particles,
such as pions and kaons, which will help us study the effect of mass shift on two-particle correlations.
An even more interesting study based on our analysis will estimate the effect of expansion on the
time-like freeze out process using the Bjorken model \cite{new_article}.
Therefore, we believe that our model may give a better description and understanding of the final
observables which are calculated using the single (and two) particle distribution functions.

\section*{ACKNOWLEDGMENTS}

The authors, L. P. Csernai, E. Moln\'ar, A. Ny\'iri and K. Tamosiunas  thank
the hospitality of the University of Cape Town, where parts of this work were done.
E. Moln\'ar, also thanks the hospitality of the Babe\c s-Bolyai University of Cluj.
\\ \indent
Enlightening discussions with Cs. Anderlik,
T. S. Bir\'o, J. Cleymans, A. Dumitru and S. Zschocke are gratefully acknowledged.

\section*{APPENDIX A}

Here we discuss the properties of the rethermalization term from eq. (\ref{first-rethermalized})
and its consequences on finishing the FO process in a finite layer.
\begin{figure}[t!]
\centering
\includegraphics[width=8.6cm, height = 2.4cm]{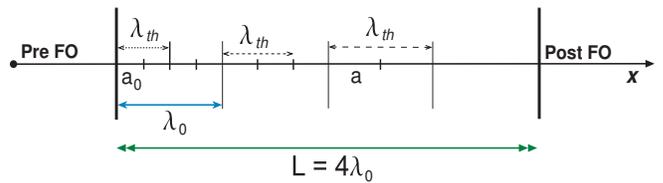}
\caption{(Color online) A schematic view of the FO process for a linear density profile.
Initially the mean free path is $\lambda_{mfp} = a_0$, the relaxation length is $\lambda_{th} = 2 \lambda_{mfp}$,
the initial characteristic length is $\lambda_0$, while the length of the FO layer is $L = 4\lambda_0$.}
\label{figure_11t}
\end{figure}
\\ \indent
From kinetic theory we know that if the following conditions:
\be
\tau_{mfp} < \tau_{th} < \tau_0 \quad \text{or}  \quad
\lambda_{mfp} < \lambda_{th} < \lambda_{0}\, ,
\ee
between the average length between the collisions, the relaxation length, and the characteristic length
are satisfied, then we can use the Boltzmann Transport Equation (BTE) for the evolution of the single
particle distribution function, $f(x,p)$.
\\ \indent
For better understanding we first assume a linear decrease of the interacting particle density during
freeze out, such as, $n(x) = (L - x)/L$, where the mean free path of interacting particles is,
$\lambda_{mfp}(x) \approx 1/n(x)$.
Using the relaxation time approximation for the FO process, for example at $x>2\lambda_0$ the density
of the interacting particles already decreased to $n(x) < n_0/2$, while the mean free path increased to
$\lambda_{mfp}(x) > a =2a_0 $, see Fig. \ref{figure_11t}.
Consequently, by the end of the FO process, the thermalization length becomes longer than the initial characteristic
length of the system.
Although, $\tau_0$ is constant (scale parameter), the characteristic FO length is actually not constant during the evolution.
Thus, using the rethermalization approximation the error we introduce within is of the order of $\tau_{th}/\tau_0$.
\\ \indent
In our model the change in the density is generally given as:
\be\label{density}
d n_{i}(x) = u_{i,\mu}(x) \, dN^{\mu}_{i}(x) \, .
\ee
This leads to an exponentially fast decrease of particle density, therefore more than $95\%$ of the interacting  matter is frozen out before $\tau_{th}\simeq \tau_0$, thus we can safely use the
relaxation time approximation in our calculations.

\section*{APPENDIX B}
The changes in the particle four current and energy momentum tensor are:
\bea \nonumber
d N^{0}_i (t) &=& - \frac{dt}{\tau_0} \frac{L}{L-t} \frac{n}{4ju\gamma}
\Bigg\{-  G_1^+(m) + \, G_1^-(m) \Bigg\} \\ \nonumber
&\buildrel m=0 \over \longrightarrow & - \frac{dt}{\tau_0} \frac{L}{L-t} n
\Bigg[\frac{(3+v^2)}{3}\,\gamma^2  \Bigg] \, ,
\eea
\bea \nonumber
d N^{x}_i (t) &=& \frac{dN^{0}_i (t)}{ju} - \\ \nonumber
&&\frac{dt}{\tau_0} \frac{L}{L-t} \frac{n}{4ju\gamma}
\Bigg\{- 2b\Big[2 K_1(a) + aK_0(a) \Big]\Bigg\} \\
&\buildrel m=0 \over \longrightarrow & - \frac{dt}{\tau_0} \frac{L}{L-t} n
\Bigg[\frac{(3+v^2)}{3v}\,\gamma^2  - \frac{1}{v}\Bigg] \, ,
\eea
\bea \nonumber
d T^{00}_i (t) &=& - \frac{dt}{\tau_0} \frac{L}{L-t} \frac{nT}{4ju\gamma}
\Bigg\{ - G_2^+(m) + \, G_2^-(m) \Bigg\} \\ \nonumber
&\buildrel m=0 \over \longrightarrow & - \frac{dt}{\tau_0} \frac{L}{L-t} nT
\Bigg[ 3(1+v^2)\,\gamma^3  \Bigg] \, ,
\eea
\bea \nonumber
d T^{0x}_i (t) &=& \frac{dT^{00}_i(t) }{ju} -
\frac{dt}{\tau_0} \frac{L}{L-t} \frac{nT}{4ju\gamma} \\ \nonumber
&&\Bigg\{- 2b^2(3 + u^2) K_2(a) - 2ab^2 K_1(a) \Bigg\} \\ \nonumber
&\buildrel m=0 \over \longrightarrow & - \frac{dt}{\tau_0} \frac{L}{L-t} nT
\Bigg[ \frac{3(1+v^2)\,\gamma^3}{v} - \frac{\gamma (3+v^2)}{v}\Bigg] \, ,
\eea
\bea \nonumber
d T^{xx}_i (t) &=& \frac{dT^{0x}_i(t)}{ju} - \frac{T}{\gamma ju} \Bigg[ dN^{x}_i(t) -
\frac{dN^{0}_i(t)}{ju} \Bigg] \\ \nonumber
&-& \frac{dt}{\tau_0} \frac{L}{L-t} \frac{nT}{4ju\gamma}
\Bigg\{-\frac{2b^2}{ju} (1 + 3u^2) K_2(a) \\ \nonumber
&-& 2juab^2 K_1(a) \Bigg\} \\ \nonumber
&\buildrel m=0 \over \longrightarrow & - \frac{dt}{\tau_0} \frac{L}{L-t} nT
\Bigg[ \frac{3(1+v^2)\,\gamma^3}{v^2}
- \frac{\gamma (3+v^2)}{v^2} \\ \nonumber
&+& \frac{1}{\gamma v^2} - \frac{(1 + 3v^2)\gamma}{v^2}\Bigg] \, ,
\eea
\bea \nonumber
d T^{yy}_i (t)&=& - \, \frac{dT^{xx}_i(t)}{2}
- \frac{dt}{\tau_0} \frac{L}{L-t} \frac{nT}{8ju\gamma} \\ \nonumber
&& \Bigg\{ - G^+_{3}(m) + G^+_{3}(m) \Bigg\} \\ \nonumber
&\buildrel m=0 \over \longrightarrow & - \frac{dt}{\tau_0} \frac{1}{2}
\Bigg[ dT^{xx}(t) + dT^{00}(t)\Bigg] \, ,
\eea
and
\bea
d T^{zz}_i (t) = d T^{yy}_i (t)\, ,
\eea
where $a = \frac{m}{T}$, $b=a\gamma$ and
$n = 4 \pi T^3 a^2 K_2(a)\, g \frac{e^{\mu/T}}{(2 \pi \hbar)^3}$ is the particle
density, while $g$ is the degeneracy factor.
Furthermore, the definition of the modified Bessel function of the second kind $K_{n}(z)$ for $n>-1$, is
\be
K_{n}(z) = \frac{2^n \, n!}{(2n)!} \, z^{-n}
\int_{z}^{\infty} dx \,e^{-x} \,(x^2 - z^2)^{n-\frac{1}{2}} \, .
\ee
The analytically not integrable functions $G_n^- (m) $ and
$G_n^+ (m) $, where $n>-2$, depend on other quantities such as $u$ and $T$.
However, in the case when, $m \rightarrow 0$, the dependence on other quantities persists and hence
we only denote the mass dependence of the functions defined as:
\bea G_n^{\pm}(m)
&=&\frac{1}{T^{n+2}}\int_{0}^{\infty} d p \, p \, \Big( \sqrt{p^2 + m^2} \Big)^n \, \\ \nonumber
&\times&\Gamma \Big(0,\frac{\gamma}{T}
\sqrt{p^2 + m^2} \pm \frac{\gamma jup}{T}\Big) \, .
\eea
The exponential integral function is defined as:
\be\label{exponential_integral}
\textrm{Ei} (z) = \int_{z}^{\infty} dx \, \frac{e^{-x}}{x} \, .
\ee


\end{document}